\documentstyle[12pt,epsfig]{article}
\voffset     = - 0.10 in
\begin{document}
\thispagestyle{empty}




\vspace*{1cm}

\begin{center}
{\Large\bf  
Longitudinal Hadronic Shower Development \\
in a Combined Calorimeter}
\end{center}

\bigskip
\bigskip

\begin{center}
{\large\bf Y.A.~Kulchitsky, M.V.~Kuzmin} 

\smallskip
{\sl Institute of Physics, National Academy of Sciences, Minsk, Belarus}
\\
\smallskip
\& \ \  
{\sl JINR, Dubna, Russia}
\bigskip

{\large\bf V.B.~Vinogradov}
\smallskip

{\it JINR, Dubna, Russia}

\end{center}

\vspace*{\fill}

\begin{abstract}
This work is devoted to  the experimental study
of the longitudinal ha\-d\-ro\-nic shower development
in the ATLAS barrel combined prototype ca\-lo\-ri\-me\-ter
consisting of the lead-liquid argon electromagnetic part and 
the iron-scintillator
hadronic part.
The results have been obtained on the basis of 
the 1996 combined test beam data which have been
taken on the H8 beam of the CERN SPS, with the pion beams 
of 10, 20, 40, 50, 80, 100, 150 and 300 GeV/c.
The degree of description of generally accepted  
Bock parameterization of the longitudinal shower  development
has been investigated.
It is shown that this parameterization does not give satisfactory description
for this combined calorimeter.
Some modification of this parameterization, in which the $e/h$ ratios of
the compartments of the combined calorimeter are used,
is suggested and compared with the experimental data.
The agreement between such parameterization and the experimental data
is demonstrated.
\end{abstract}

\newpage

\section{Introduction}
\hspace{6mm}
One of the important questions of hadron calorimetry is the question
of the longitudinal development of hadronic showers.
This question is especially important for a combined calorimeter.
This work is devoted to the study of the longitudinal hadronic 
shower development in the ATLAS barrel combined prototype calorimeter
\cite{atcol94,TILECAL96,LARG96,ajaltouni97,cobal98}.   

This work has been performed on the basis of the 1996 combined test beam data 
\cite{cobal98}.
Data were taken on the H8 beam of the CERN SPS with the pion beams 
of 10, 20, 40, 50, 80, 100, 150 and 300 GeV/c.

\section{Combined Calorimeter}
\begin{figure}[tbh]
\begin{center}
\mbox{\epsfig{figure=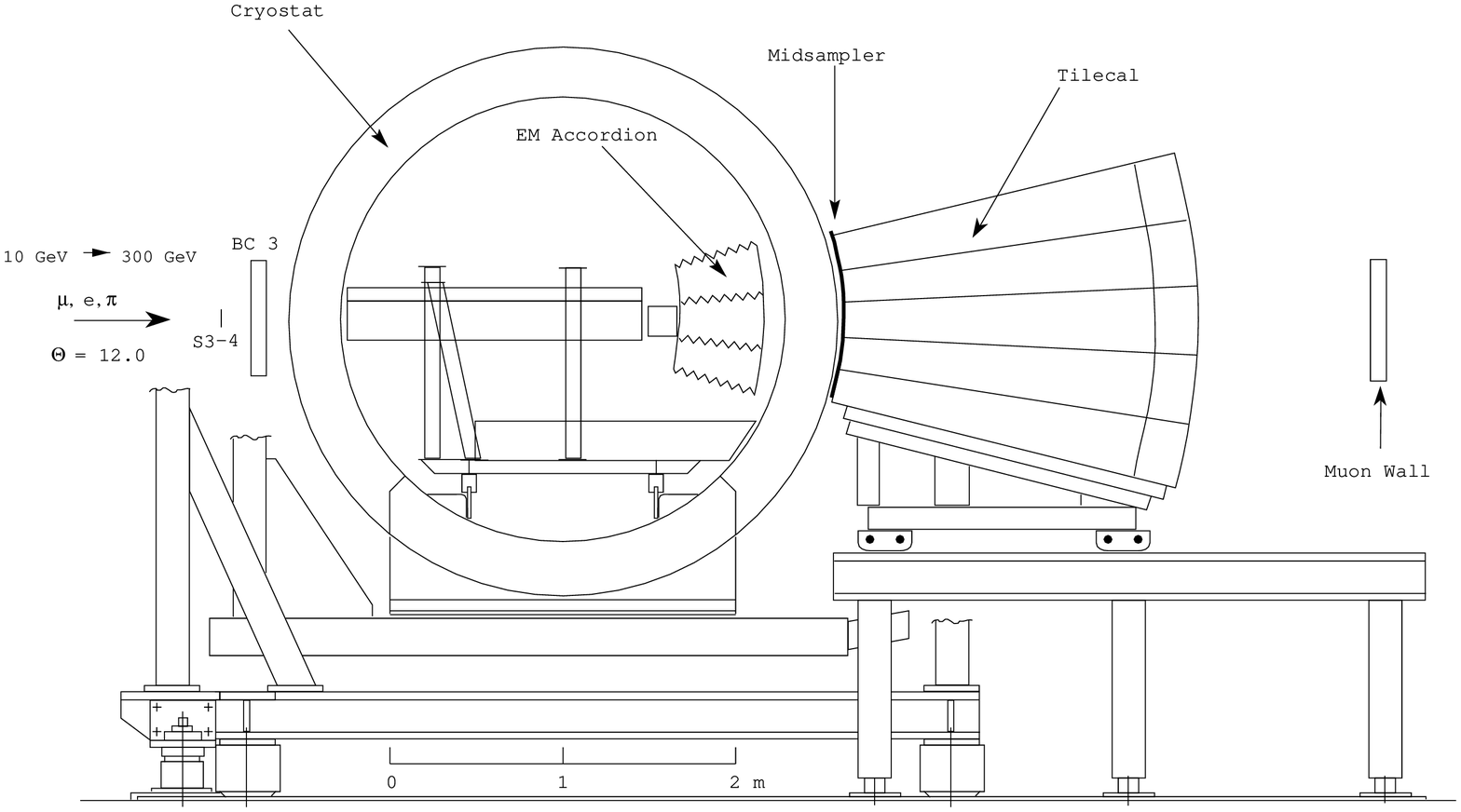,width=0.95\textwidth,height=0.4\textheight}} 
\end{center}
 \caption{ 
   Test beam setup for the combined LAr and Tile calorimeters run.}
\label{fv1-01}
\end{figure}

\hspace{6mm}
The future ATLAS experiment 
\cite{atcol94} 
will include in the central (``barrel'') region a calorimeter system composed
of two separate units: 
the liquid argon electromagnetic calorimeter (LAr)
\cite{LARG96}
and the tile iron-scintillating hadronic calorimeter (Tile) 
\cite{TILECAL96}. 
For detailed understanding of performance of the future ATLAS
combined calorimeter the combined calorimeter prototype setup
has been made consisting of the LAr electromagnetic calorimeter prototype
inside the cryostat and downstream the Tile calorimeter prototype
as shown in Fig.~\ref{fv1-01}. 
The dead material upstream of the LAr calorimeter was about 
$0.1\ \lambda_{\pi}$ 
and the one between the two calorimeters was about 
$0.3\ \lambda_{\pi}$.
The two calorimeters have been placed with their central axes at an angle 
to the beam of $12^\circ$.
At this angle the two calorimeters have an active thickness of 
$8.6\ \lambda_{\pi}$.
Between the active part of the LAr and the Tile detectors
a layer of scintillator was installed,  called the midsampler. 
The midsampler consists of five scintillators,  $20\times100$\ cm$^2$ each, 
fastened directly to the front face of the Tile modules.
The scintillator is 1 cm thick.
Beam quality and geometry were monitored with a set 
of beam wire chambers BC1,  BC2,  BC3 and trigger hodoscopes placed upstream 
of the LAr cryostat.
To detect punchthrough particles and to measure the effect of longitudinal 
leakage a ``muon wall'' consisting of 10 scintillator counters (each 2 cm 
thick) was located behind the calorimeters at a distance of about 1 metre.

\subsection{Electromagnetic Calorimeter}
\hspace{6mm}
The electromagnetic LAr calorimeter  prototype consists of a stack of three 
azimuthal modules, each one spanning $9^\circ$ in azimuth and extending over 
2 m along the Z direction.
The calorimeter structure is defined by 2.2 mm thick steel-plated lead 
absorbers,  folded to an accordion shape and separated by 3.8 mm gaps,  
filled with liquid argon.
The signals are collected by Kapton electrodes located in the gaps.
The calorimeter extends from an inner radius of 131.5 cm to an outer radius 
of 182.6 cm, representing (at $\eta = 0$) a total of 25 radiation lengths 
($X_0$),  or 1.22 interaction lengths ($\lambda_I$) for protons.
The calorimeter is longitudinally segmented into three compartments of
$9\ X_0$,  $9\ X_0$ and $7\ X_0$, respectively.
More details about this prototype can be found in 
\cite{atcol94,ccARGON}.
In front of the EM calorimeter a presampler was mounted. 
The active depth of liquid argon in the presampler is 10 mm and
the strip spacing is 3.9 mm.
The cryostat has a cylindrical form
with 2 m internal diameter,  filled with liquid argon,  and
is made out of a  8 mm thick inner stainless-steel vessel, 
isolated by 30 cm of low-density foam (Rohacell),  itself protected by a
1.2 mm thick aluminum outer wall.

\subsection{Hadronic Calorimeter}
\hspace{6mm}
The  hadronic Tile calorimeter is a sampling device using
steel as the absorber and scintillating tiles as the active material 
\cite{TILECAL96}.
The innovative feature of the design is the orientation of the tiles
which are placed in planes perpendicular to the Z direction 
\cite{gild91}.
For a better sampling homogeneity the 3 mm thick scintillators are staggered 
in the radial direction.
The tiles are separated along  Z  by 14 mm of steel,  giving a
steel/scintillator volume ratio of 4.7.
Wavelength shifting fibers (WLS) running radially collect light from the
tiles at both of their open edges.
The hadron calorimeter prototype consists of an azimuthal stack of five 
modules.
Each module covers $2\pi/64$ in azimuth and extends 1 m along the Z direction,
such that the front face covers $100\times20$\ cm$^2$. 
Read-out cells are defined by grouping together a bundle of fibers
into one photomultiplier (PMT).
Each of the 100 cells is read out by two PMTs 
with $\Delta \phi = 2\pi/64 \approx 0.1$, while the segmentation 
along the Z axis is made by grouping fibers into read-out cells spanning
$\Delta Z = 20$ cm 
($\Delta \eta \approx 0.1$).
Each module is read out in four longitudinal segments (corresponding to about 
1.5,  2,  2.5 and 3 $\lambda_I$ at $\eta = 0$).
More details of this prototype can be found in
\cite{atcol94,ariztizabal94}.

\section{Event Selection}
\hspace{6mm}
We applied some  cuts similar to 
\cite{cobal98} 
to eliminate the nonsingle track pion events, the  beam halo,  the events 
with an interaction before LAr calorimeter and muon events.
The set of cuts applied is the following:
\begin{itemize}
\item
single-track pion events were selected off-line by requiring the pulse height 
of the beam scintillation counters and the energy released in the presampler 
of the electromagnetic calorimeter to be compatible with that of a single 
particle;
\item   
beam halo events were removed with appropriate cuts on the horizontal 
and  vertical positions of the incoming track impact point
and the space angle with respect to the beam axis
as measured with the two beam chambers;
\item
a cut on the total energy rejects incoming muon.
\end{itemize}

\section{Energy Reconstruction} 
\hspace{6mm}
To reconstruct the hadron energy in longitudinal segments the $e/h$ method
of the energy reconstruction, suggested in \cite{kulchitsky99-2}, 
has been used.
In this method the energy of hadrons in a combined calorimeter is determined 
by the following formula:
\begin{eqnarray}
	E  = 
		c_{LAr}  \cdot (e/\pi)_{LAr}  \cdot R_{LAr}  + 
		c_{Tile} \cdot (e/\pi)_{Tile} \cdot R_{Tile} + 
		E_{dm} \ ,
\label{ev7}
\end{eqnarray}
Here $R_{LAr}$ ($R_{Tile}$) is the LAr (Tile) calorimeter response,
$c_{LAr}  = 1 / e_{LAr}$,  $c_{Tile} = 1 / e_{Tile}$, $e_{LAr}$ and $e_{Tile}$
are the electron calibration constants for LAr and Tile calorimeters, 
$e/\pi$ ratios are equal to
\begin{eqnarray}
	\Bigl( \frac{e}{\pi} \Bigr)_{LAr} 
		= \frac{( e/h)_{LAr}}{1+((e/h)_{LAr}-1)f_{\pi^0, LAr}} \  .
\label{ev5}
\end{eqnarray} 
\begin{eqnarray}
	\Bigl(\frac{e}{\pi} \Bigr)_{Tile} = 
		\frac{( e/h)_{Tile}}{1+((e/h)_{Tile}-1)f_{\pi^0, Tile}} \ .
\label{ev6}
\end{eqnarray}  
This method uses only the known $e/h$ ratios and the electron calibration 
constants  and does not require the previous determination of any parameters 
by a minimization technique.
The value of the $(e/h)_{LAr}$ ratio of the electromagnetic compartment
has been obtained in \cite{kulchitsky99-1} and equal to 
$(e/h)_{LAr} = 1.77\pm0.02$.
This value agrees with with estimation of $> 1.7$ obtained in 
\cite{wigmans91}.
  
For Tile calorimeter the value of $(e/h)_{Tile} = 1.30\pm0.04$ is used 
\cite{budagov95,kulchitsky99-12}.

The fraction of the shower energy going into the electromagnetic channel 
for LAr compartment is
\begin{eqnarray}
 	f_{\pi^0, LAr} =  
			0.11 \cdot ln{(E_{beam})} \ .
\label{ev4}
\end{eqnarray}

The electromagnetic fraction in the Tile calorimeter, which samples the 
final part of shower, is equal to the one for shower with energy $E_{Tile}$:
\begin{eqnarray}
	f_{\pi^0, Tile}= 
			0.11 \cdot ln{(E_{Tile})} \ ,
\label{ev3}
\end{eqnarray}
where
\begin{eqnarray}
	E_{Tile}  = 
			c_{Tile} \cdot (e/\pi)_{Tile} \cdot R_{Tile} \ .
\label{ev16}
\end{eqnarray}

\begin{figure*}[tbph]
\begin{center}   
\begin{tabular}{c}
\mbox{\epsfig{figure=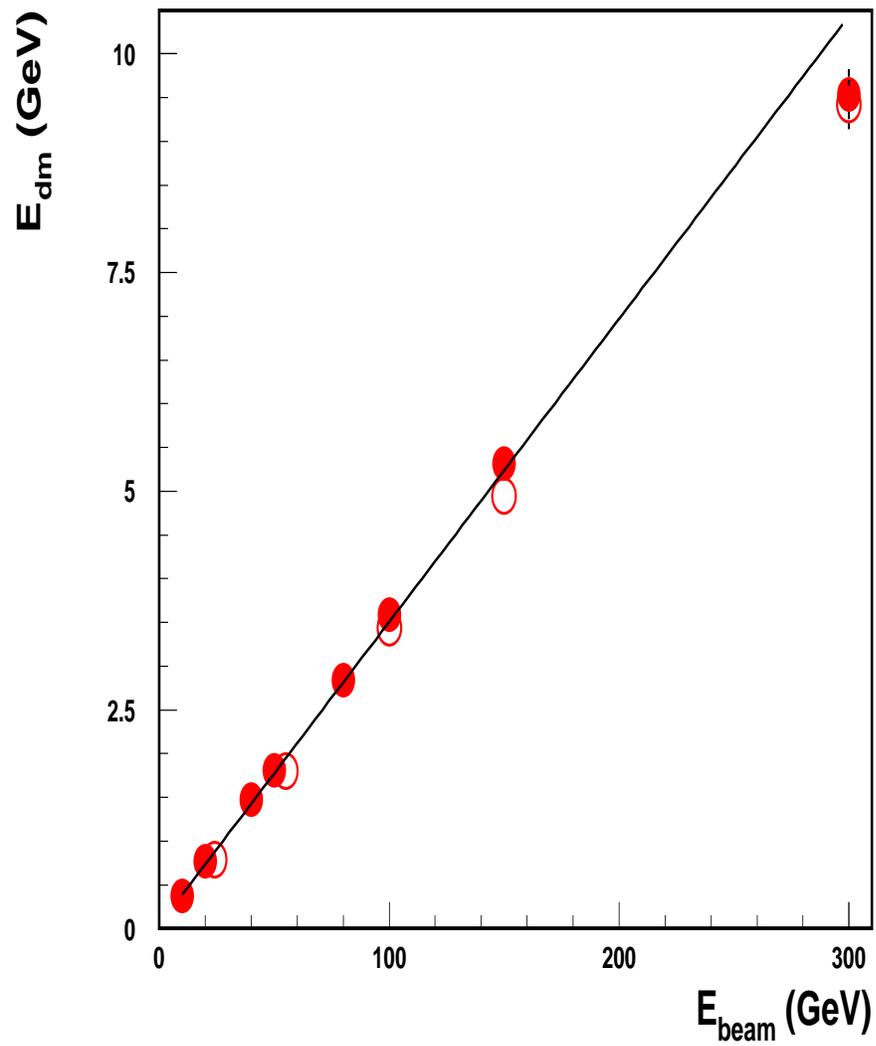,width=0.95\textwidth,height=0.95\textheight}} 
\\
\end{tabular}
\vspace*{-20 mm}
\caption{
	The comparison between the Monte Carlo simulation (open circles)
	and the calculated values (solid circles).
	}
\label{fmc}
\end{center}
\end{figure*}

Special attention has been devoted to understanding of the energy loss in 
the dead material placed between the active part of the LAr and the Tile 
detectors.
The term  $E_{dm}$ accounts for this.
This term is taken to be proportional to the geometrical mean 
of the energy released in the last electromagnetic compartment
($E_{LAr, 3}$) and the first hadronic compartment ($E_{Tile, 1}$)
\begin{eqnarray}
\label{ev19}
	E_{dm} = 
			c_{dm} \cdot \sqrt{E_{LAr, 3} \cdot E_{Tile, 1}} 
\end{eqnarray}
similar to 
\cite{cobal98}.
We used the value of  $c_{dm} = 0.31$.
This value has been obtained on the basis of the results of the Monte Carlo 
simulation \cite{efthymiopoulos}. 
These Monte Carlo (Fluka) results (open circles) are shown in  
Fig.\ \ref{fmc} 
together with the values (solid circles) obtained by using the expression 
(\ref{ev19}).
The good agreement is observed.
Also the linear behaviour as a function of the beam energy is demonstrated. 
The mean energy loss is  equal to about $3.7\pm0.4\%$ (spread).

\begin{figure*}[tbph]
\begin{center}   
\begin{tabular}{c}
\mbox{\epsfig{figure=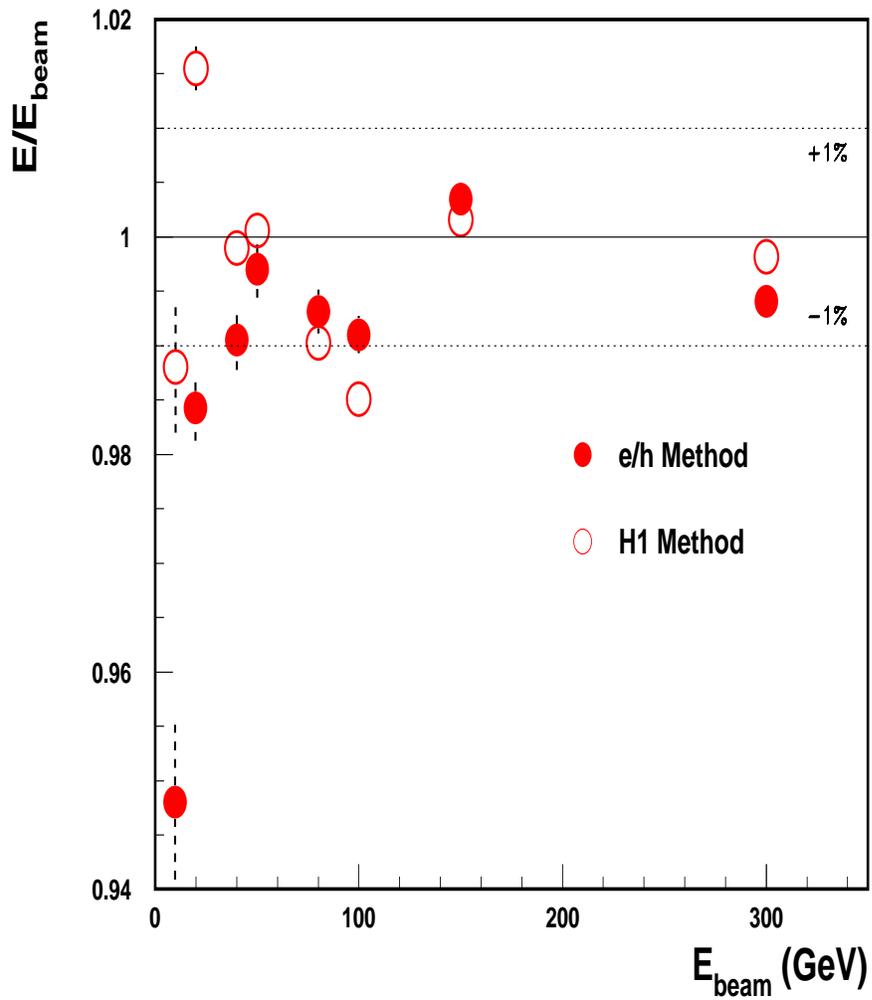,width=0.95\textwidth,height=0.9\textheight}} 
\\
\end{tabular}
\vspace*{-20 mm}
\caption{
	The demonstration of the correctness of the reconstruction
	of the mean values of energies by the $e / h$ method (black circles):
	the energy linearity, $E_{rec}/E_{beam}$ as a function of 
	the beam energy. It is also shown the same for the cells weighting 
	$H1$ method (open circles).
	}
\label{f03}
\end{center}
\end{figure*}

The $e/h$ method  \cite{kulchitsky99-2}
has been tested on the basis of the 1996 test beam data
of the ATLAS combined prototype calorimeter and demonstrated the correctness 
of the reconstruction of the mean values of energies 
as shown in Fig.~\ref{f03}.
As can be seen the deviation from linearity for the $e/h$ method is about 
$1\%$. 

\begin{figure*}[tbph]    
 \begin{center}
        \begin{tabular}{c}
\mbox{\epsfig{figure=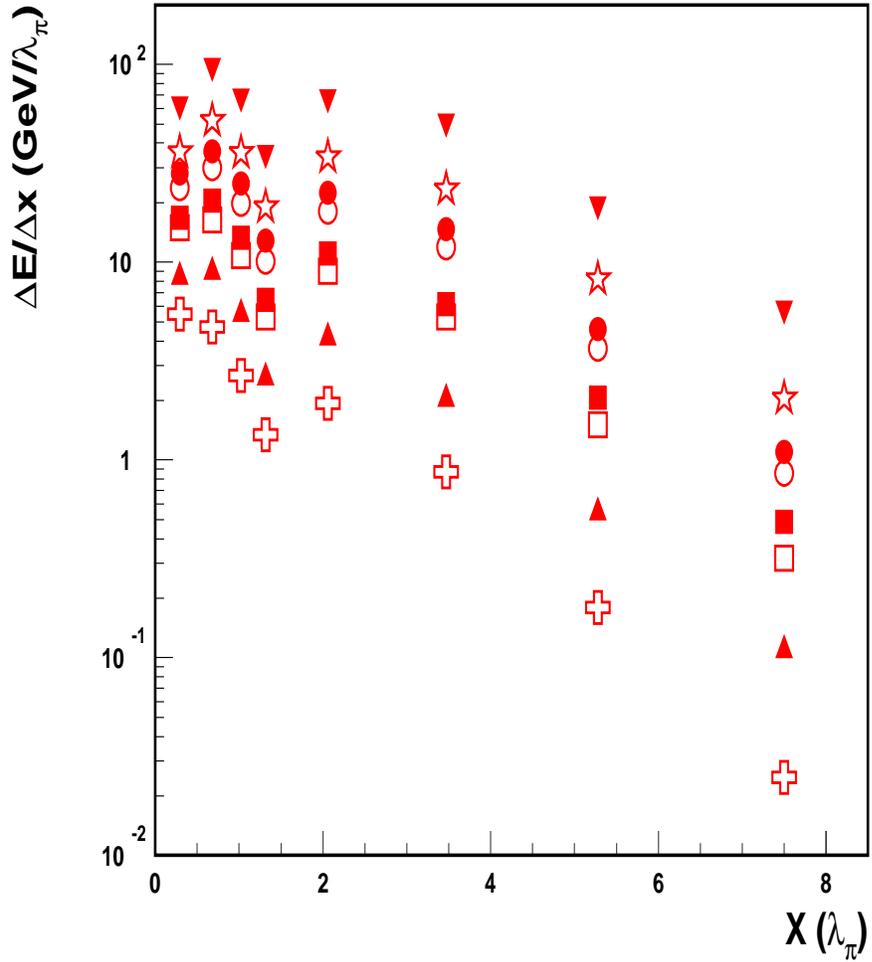,width=0.95\textwidth,height=0.88\textheight}}
        \\
        \end{tabular}
\vspace*{-20 mm}
\caption{
	The differential longitudinal energy depositions
	in each longitudinal sampling for
	10 GeV (crosses), 
	20 GeV (black top triangles),
        40 GeV (open squares),
        50 GeV (black squares),
        80 GeV (open circles),
        100 GeV (black circles),
        150 GeV (stars) and
        300 GeV (black bottom triangles) 
        energies as a function of the longitudinal coordinate $x$ in units
        $\lambda_{\pi}$ for combined calorimeter.
	}
\label{fv6}
     \end{center}
\end{figure*}

We used this energy reconstruction method and obtained the energy depositions,
$E_i$, in each longitudinal sampling with the thickness of $\Delta x_i$ in 
units $\lambda_{\pi}$.
We transformed these depositions into the differential energy depositions
using the formula:
\begin{eqnarray}
	(\Delta E/ \Delta x)_i = E_i / \Delta x_i \ .
\label{edep}
\end{eqnarray}
Table \ref{T1} and Fig.~\ref{fv6} 
show the differential energy depositions as a function of the 
longitudinal coordinate $x$ for 10 GeV (crosses), 20 GeV (black top triangles),
40 GeV (open squares), 50 GeV (black squares), 80 GeV (open circles), 
100 GeV (black circles), 150 GeV (stars), 300 GeV (black bottom triangles) 
energies.
Some interesting special features are observed:
the maximum in the region of the LAr calorimeter, then the local minimum in 
the point corresponding to the energy losses in the dead material,
then the local maximum again.        

\section{Longitudinal Shower Development}
\hspace{6mm}
The next important question is understanding and description
of these experimental data.

There is the well known parameterization of the longitudinal hadronic shower 
development from the shower origin suggested in
\cite{bock81} 
\begin{eqnarray}
        \frac{dE_{s} (x)}{dx} =
                N\
                \Biggl\{
                w\ \biggl( \frac{x}{X_0} \biggr)^{a-1}\
                e^{- b \frac{x}{X_0}}\
                + \
                (1-w)\
                \biggl( \frac{x}{\lambda_I} \biggr)^{a-1}\
                e^{- d \frac{x}{\lambda_I}}
                \Biggr\} \ ,
\label{elong00}
\end{eqnarray}
where $X_0$ is the radiation length, $\lambda_I$ is the interaction length,
$a,\ b,\ d,\ w$ are parameters, $N$ is the normalization factor,
$a = 0.6165 + 0.3193\ lnE$, $b = 0.2198$, $d = 0.9099 - 0.0237\ lnE$,
$\omega = 0.4634$.

This parameterization is from the shower origin.
But our data are from the calorimeter face and due to the unsufficient 
longitudinal segmentation can not be transformed to the shower origin.
Therefore, we used the analytical representation of the hadronic shower
longitudinal development from the calorimeter face 
\cite{kulchitsky98}.
This representation is a result of the integration of the longitudinal 
profile from the shower origin over the shower position:
\begin{eqnarray}
        \frac{dE (x)}{dx} = 
        \int \limits_{0}^{x} 
        \frac{dE_{s}(x-x_v)}{dx}\ e^{- \frac{x_v}{\lambda_I}}\ dx_v \ ,
\label{elong01}
\end{eqnarray}
where $x_v$ is a coordinate of the shower vertex.
This representation has the following form:
\begin{eqnarray}
        \frac{dE (x)}{dx} & = &
                N\
                \Biggl\{
                \frac{w X_0}{a}
                \biggl( \frac{x}{X_0} \biggr)^a
                e^{- b \frac{x}{X_0}}
                {}_1F_1 \biggl(1,a+1,
                \biggl(b - \frac{X_0}{\lambda_I} \biggr) \frac{x}{X_0}
                \biggr)
                \nonumber \\
                & & + \
                \frac{(1 - w) \lambda_I}{a}
                \biggl( \frac{x}{\lambda_I} \biggr)^a
                e^{- d \frac{x}{\lambda_I}}
                {}_1F_1 \biggl(1,a+1,
                \bigl( d -1 \bigr) \frac{x}{\lambda_I} \biggr)
                \Biggr\} .
\label{elong03}
\end{eqnarray}
Here ${}_1F_1(\alpha,\beta,z)$ is the confluent hypergeometric function  and
$N$ is the normalisation factor which equal to
\begin{eqnarray}
        N =     
                \frac{E_{beam}}{\lambda_I\ \Gamma (a)\ 
                \bigl(w\ X_0\ b^{-a} + (1-w)\ \lambda_I\ d^{-a} \bigr)} \ .  
\label{elong04}
\end{eqnarray} 

\begin{figure*}[tbph]  
   \begin{center}
        \begin{tabular}{c}
\mbox{\epsfig{figure=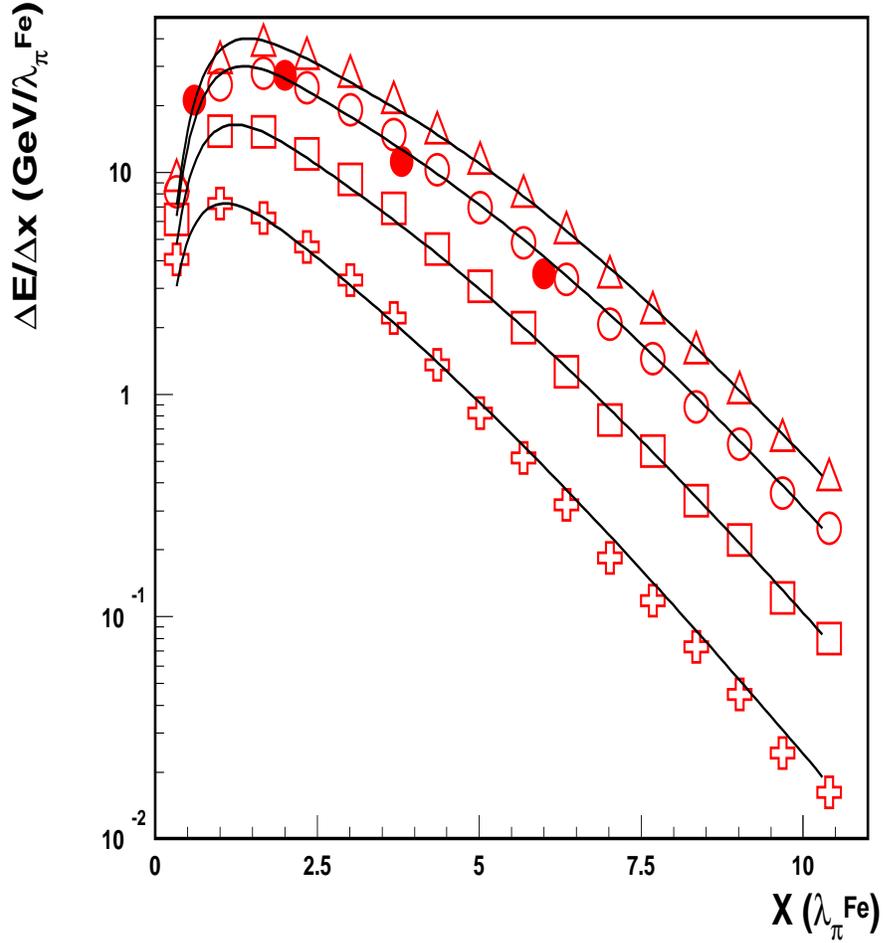,width=0.95\textwidth,height=0.85\textheight}} 
        \\
        \end{tabular}
\vspace*{-20 mm}
\caption{
	The demonstration of  the correctness of the formula (\ref{elong03}).
	The comparison between the calculations by this formula and 
	the experimental data at 20 GeV (crosses),
        50 GeV (squares), 100 GeV (open circles), 
	140 GeV (triangles) energies for conventional iron-scintillator 
        calorimeter \cite{hughes90} and at 100 GeV (black circles)
        for the Tile calorimeter as a function of the longitudinal coordinate 
	$x$ in units  $\lambda_{\pi , Fe}$
	}.      
\label{fv1}
     \end{center}
\end{figure*}

Fig.\ \ref{fv1} demonstrates the correctness of the above formula. 
The calculations by this formula  are compared with the experimental data at
20 GeV (crosses), 50 GeV (squares), 100 GeV (open circles), 
140 GeV (triangles) energies for conventional iron-scintillator calorimeter 
\cite{hughes90} 
and at 100 GeV (black circles) for the Tile calorimeter as a function of the 
longitudinal coordinate $x$ in units $\lambda_{\pi , Fe}$. 
The good agreement is observed.
Note that formula (\ref{elong00}) is given for a calorimeter 
characterizing by 
the certain $X_0$ and $\lambda_I$ values.

\begin{figure*}[tbph]     
\begin{center}
        \begin{tabular}{c}
\mbox{\epsfig{figure=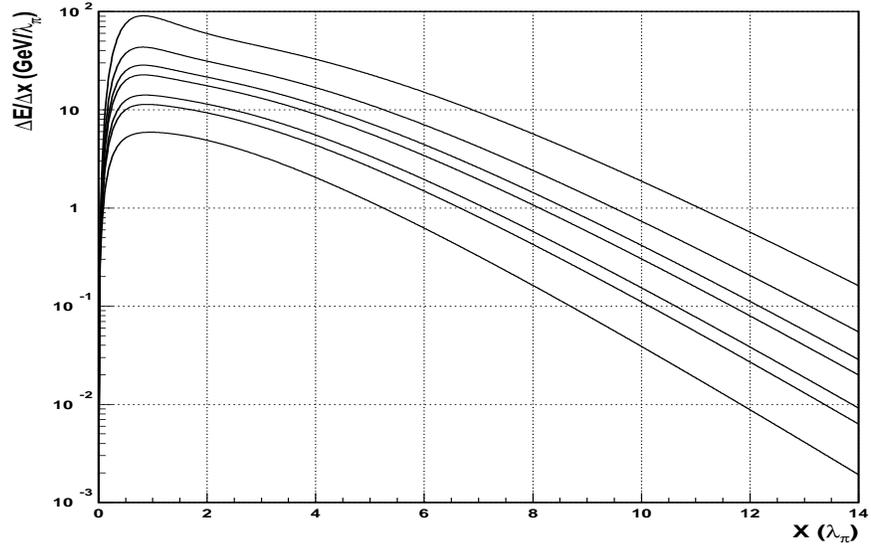,width=0.95\textwidth,height=0.43\textheight}}
        \\
\mbox{\epsfig{figure=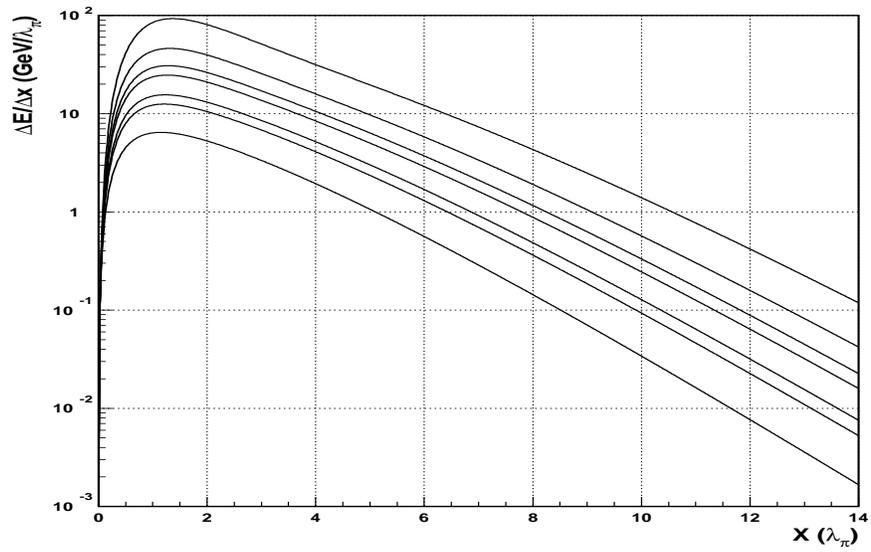,width=0.95\textwidth,height=0.43\textheight}}
        \\
        \end{tabular}
\vspace*{-5 mm}
\caption{
	The calculated differential longitudinal energy deposition, $dE/dx$,
        as a function of the longitudinal coordinate $x$ in units
        $\lambda_{\pi}$ for the LAr (top) and Tile (bottom)
 	calorimeters at $20 - 300\ GeV$.
	}
\label{fv2-1}
     \end{center}
\end{figure*}

\begin{figure*}[tbph]    
 \begin{center}
        \begin{tabular}{c}
\mbox{\epsfig{figure=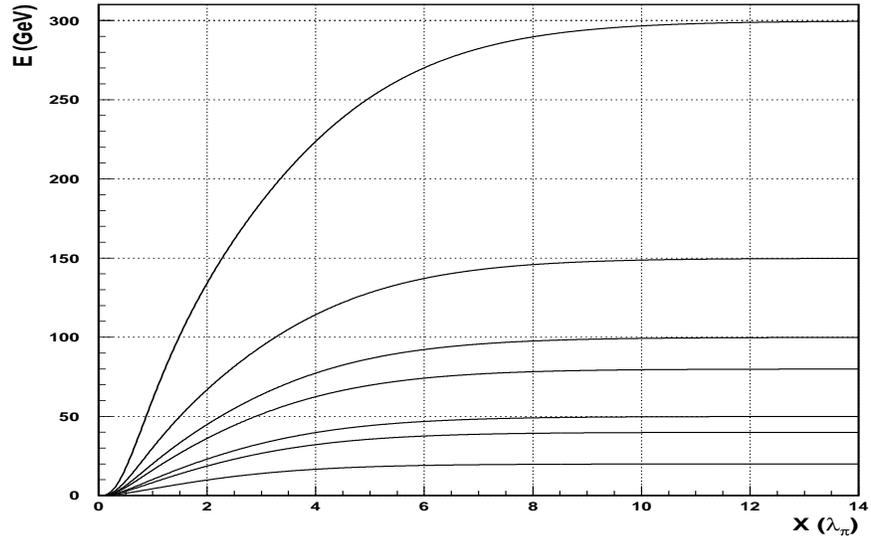,width=0.95\textwidth,height=0.43\textheight}}
\\
\mbox{\epsfig{figure=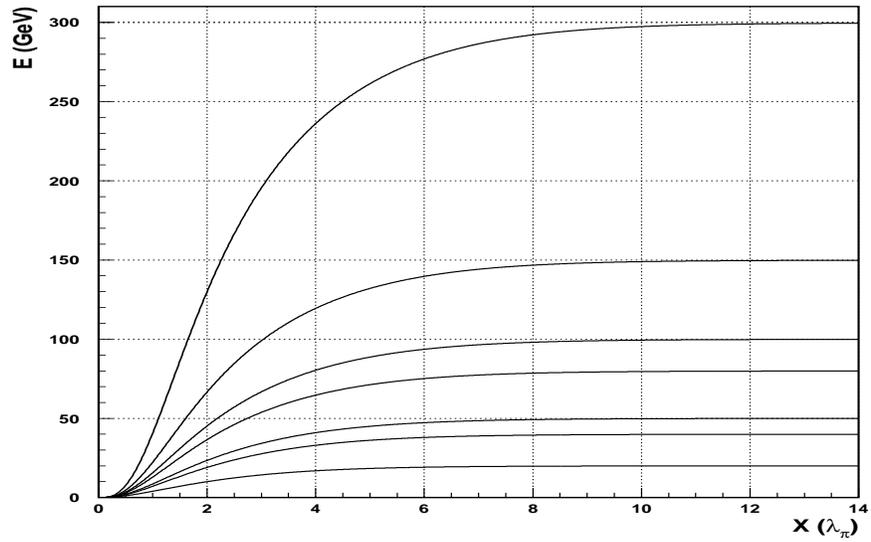,width=0.95\textwidth,height=0.43\textheight}}
        \\
        \end{tabular}
\vspace*{-5 mm}
\caption{
	The integrated longitudinal energy depositions, 
	$E(x)$ $=$  $\int dE/dx dx$, as a function of
        the longitudinal coordinate $x$ in units of
        $\lambda_{\pi}$ for the LAr (top) and Tile (bottom) calorimeters
        at 20 -- 300 $GeV$.
	} 
\label{fv3-2}
     \end{center}
\end{figure*}

We have found in literature no algorithm for the hadronic shower development 
in a combined calorimeter. 
We suggest the following algorithm of combination of the LAr and Tile curves 
of the differential longitudinal energy deposition $dE/dx$ 
(Fig.\ \ref{fv2-1}).  
The first part of the combined curve is the beginning of the LAr curve, 
the second part is the Tile curve.
At first a hadronic shower  develops  in the LAr calorimeter to the boundary 
value $x_{LAr}$.
This value corresponds to certain integrated measured energy $E_{LAr}$.
Then using the corresponding integrated Tile curve, 
$E(x) = \int_0^x (dE/dx)dx$, (Fig.\ \ref{fv3-2})
we find the point $x_{Tile}$ in which the energy is equal to 
$E_{Tile}(x_{Tile}) = E_{LAr} + E_{dm}$.
From this point a shower continues to develop in the Tile calorimeter.
In principle, instead of the measured value of $E_{LAr}$ 
one can use the calculated value of
$E_{LAr} = \int_0^{x_{LAr}} (dE/dx) dx$
obtained from the integrated LAr curve
(Fig.\ \ref{fv3-2}).
 
\begin{figure*}[tbph]    
 \begin{center}
        \begin{tabular}{c}
\mbox{\epsfig{figure=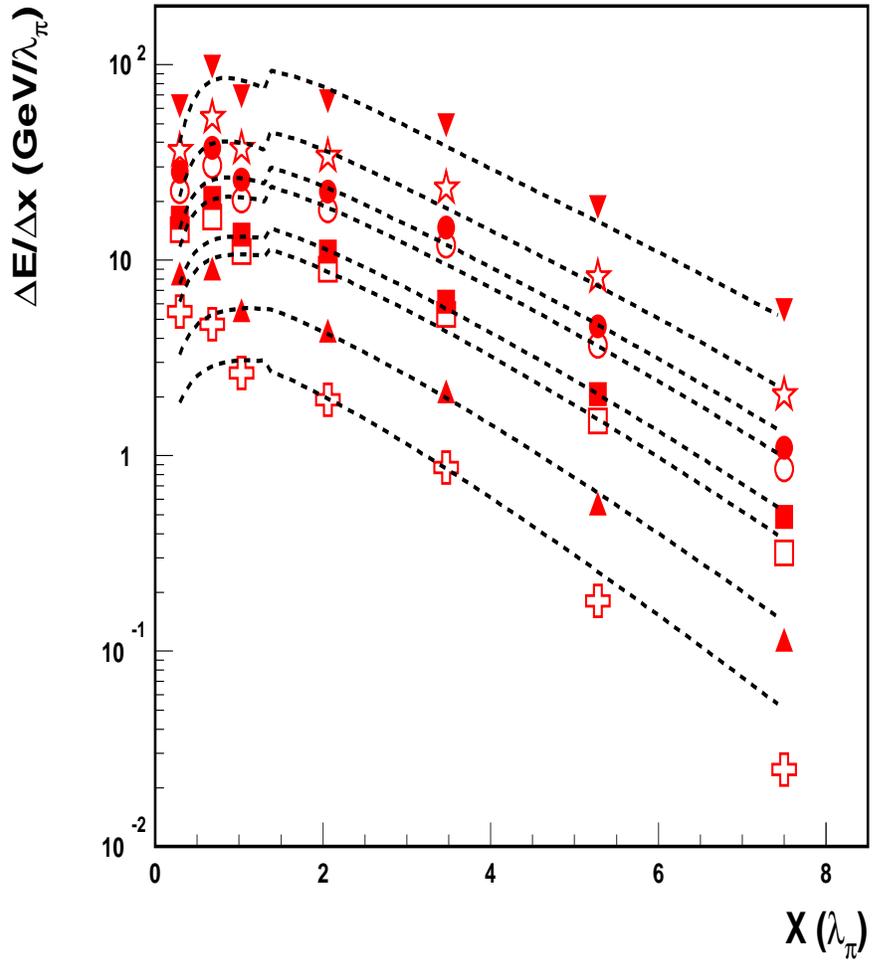,width=0.95\textwidth,height=0.87\textheight}}
        \\
        \end{tabular}
\vspace*{-20 mm}
\caption{
	The comparison between the experimental differential energy 
	depositions at 10 GeV (crosses), 20 GeV (black top triangles),
        40 GeV (open squares), 50 GeV (black squares), 80 GeV (open circles),
        100 GeV (black circles), 150 GeV (stars), 300 GeV (black bottom 
	triangles) and the calculated curves
        as a function of the longitudinal coordinate $x$ in units
        $\lambda_{\pi}$.
	}
\label{fv6-1}
     \end{center}
\end{figure*}

In this  way we obtained the combined curves. 
Fig.\ \ref{fv6-1} shows a comparison between the experimental differential 
energy depositions at 10 GeV (crosses), 20 GeV (black top triangles), 
40 GeV (open squares), 50 GeV (black squares), 80 GeV (open circles), 
100 GeV (black circles), 150 GeV (stars), 300 GeV (black bottom triangles) 
as a function of the longitudinal coordinate $x$ in units $\lambda_{\pi}$.
It can be seen that there is a significant disagreement between 
the experimental data and the combined curves in the region of the LAr 
calorimeter and especially at low energies.

We attempted to understand this disagreement.
We considered the experimental database used by Bock et al.\ \cite{bock81}
for their parametrisation.
It turns out that:
\begin{itemize}
\item
The parameters given in \cite{bock81} parameterization have been obtained 
by using  the data  from three iron-scintillator calorimeters and one 
lead-scintillator calorimeter:
\begin{itemize}
\item
CERN, WA1, $Fe$ (50 mm) + $Sc$ (6 mm), pions at 15, 50, 140 GeV, 
$e/h\approx 1.3$;
\item
FNAL, 379, $Fe$ (38.4 -- 51.2 mm) + $Sc$ (6.4 mm), pions at 375 and 400 GeV, 
$e/h\approx 1.3$;
\item
CERN, the combined UA1 calorimeter, the electromagnetic part: 
$Pb$ (2 -- 3 mm) + $Sc$ (1.5 -- 2 mm), pions at 10, 20, 40, 60 GeV, 
$e/h\approx 1.1$ and the hadronic part: $Fe$ (50 mm) + $Sc$ (10 mm), 
$e/h\approx 1.3$. 
\end{itemize}
As to the used iron-scintillator calorimeters there is sufficient number of 
experimental points in our (10 -- 300 GeV) beam energy range.
At the same time the situation for the lead-scintillator calorimeter is 
quite different.
There is a very limited number of points and the energy range of
(10 -- 60 GeV ) is essentially lower than used in our work.
\item 
The $e/h$ ratio of the LAr calorimeter is $\approx 1.6$ times greater 
than their calorimeter.
\item
This parameterization does not include such essential
feature of a calorimeter as the $e/h$ ratio.
\end{itemize}

\begin{figure*}[tbph]   
  \begin{center}
        \begin{tabular}{c}
\mbox{\epsfig{figure=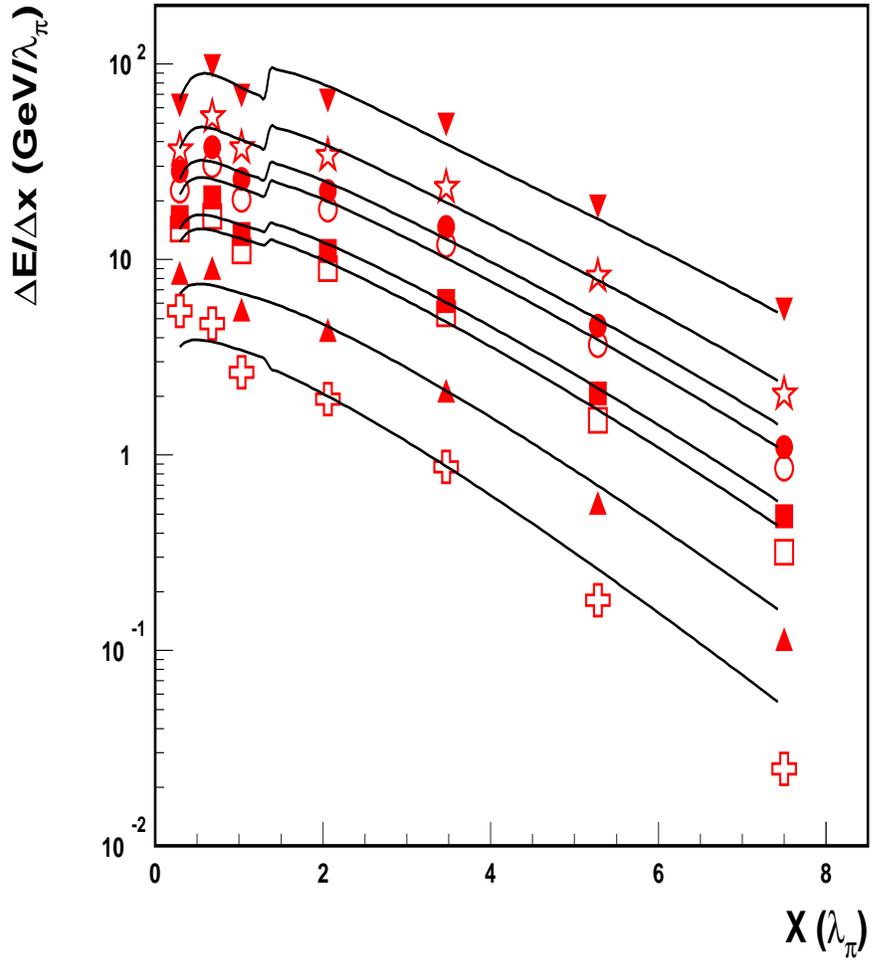,width=0.95\textwidth,height=0.87\textheight}}
        \\
        \end{tabular}
\vspace*{-20 mm}
\caption{
	The experimental differential longitudinal energy depositions at
	10 GeV (crosses), 20 GeV (black top triangles), 40 GeV (open squares),
        50 GeV (black squares), 80 GeV (open circles), 100 GeV (black circles),
        150 GeV (stars), 300 GeV (black bottom triangles)
        energies as a function of the longitudinal coordinate $x$ in units
        $\lambda_{\pi}$ for the combined calorimeter and the results 
        of the description
	by the modified parameterization.
	}
\label{fv6-2}
     \end{center}
\end{figure*}

We attempted to improve the description and tried several modifications 
and adjustments of some parameters of this parameterization.
It turned  out that the changes of two parameters $b$ and $w$
in the formula (\ref{elong03}) in such a way
$b_{Bock}	= 0.22$,
$b_{new}	= 0.34 = b_{Bock} \cdot (e/h)_{new} / (e/h)_{Bock}$,
$w_{Bock}	= 0.4634$,
$w_{new}	= 0.6 \cdot K$,
where $K$ factor is
\begin{eqnarray}
	K = 	\biggl( \frac{e}{\pi} \biggr)_{new} \biggr/
    		\biggl( \frac{e}{\pi} \biggr)_{Bock}\ \ ,
\label{elong001}
\end{eqnarray}
\begin{eqnarray}
	\frac{e}{\pi} 	= \frac{e/h}{1+(e/h - 1)f_{\pi^0}} 
\label{ev5-1}
\end{eqnarray}
made it possible to obtain the reasonable description of the experimental data.
This is shown in Fig.\ \ref{fv6-2} in which the experimental differential 
longitudinal energy depositions at 10 GeV (crosses), 20 GeV (black top 
triangles), 40 GeV (open squares), 50 GeV (black squares), 80 GeV (open 
circles), 100 GeV (black circles), 150 GeV (stars), 300 GeV (black bottom 
triangles) energies as a function of the longitudinal coordinate $x$ in units 
$\lambda_{\pi}$ for the combined calorimeter and the results of 
the description by the modified parameterization are compared.
There is a reasonable agreement (probability of description is more than
$5\%$) between the experimental data and the curves taking into account 
uncertainties in the parametrization function \cite{bock81}.
In such case the Bock parameterization is the private case for some fixed 
the $e/h$ ratio.

\begin{figure*}[tbph]
\begin{center}   
\begin{tabular}{c}
\epsfig{figure=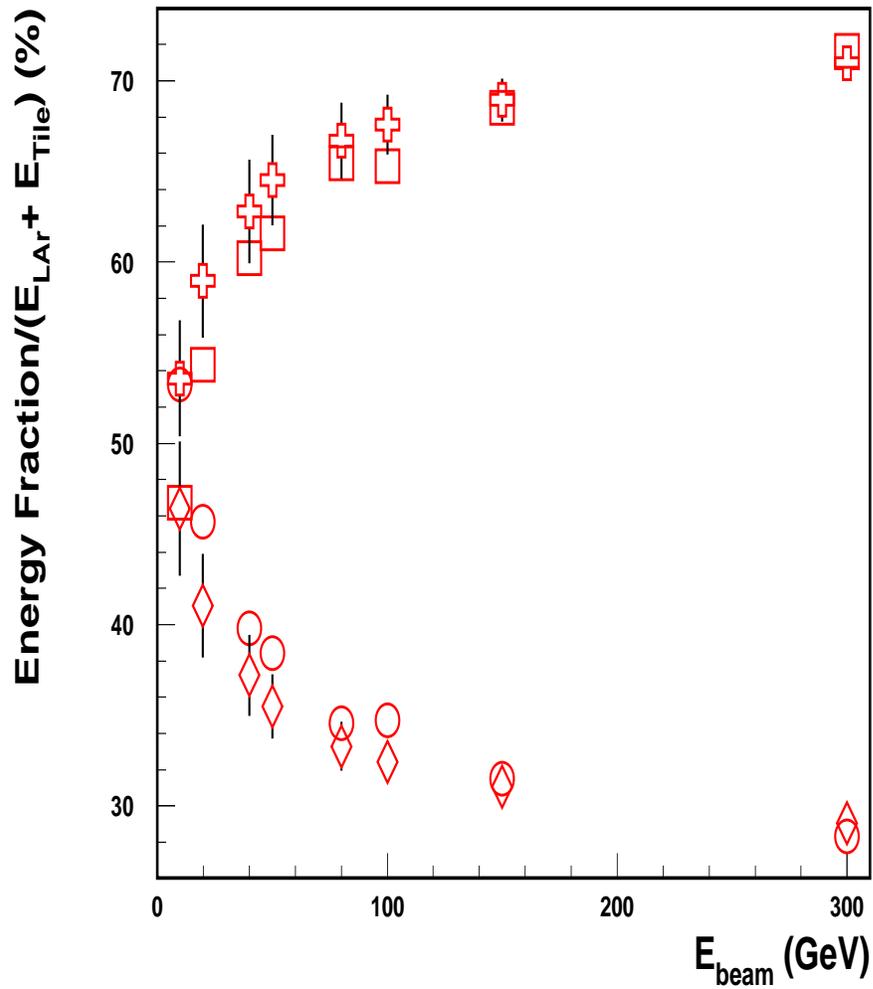,width=0.95\textwidth,height=0.9\textheight} 
\\
\end{tabular}
\end{center}
\vspace*{-20 mm}
       \caption{
         Energy deposition (percentage) in the LAr and Tile 
	 calorimeters at different beam energies.
         The circles (squares) are the measured energy depositions
	 in the LAr (Tile) calorimeter, the diamonds (crosses) are the 
	 calculated energy depositions in the ones.
	}
       \label{f04-a}
\end{figure*}

The obtained parameterization has some additional applications.
For example, this formula may be used for an estimate of the energy
deposition in various parts of a combined calorimeter.
This demonstrates in Fig.\ \ref{f04-a} in which the measured and calculated 
relative values of the energy deposition in the LAr and Tile calorimeters are
presented.
The relative energy deposition in the LAr calorimeter decreases from 
about 50\% at 10 $GeV$ to 30\% at 300 $GeV$.
On the contrary, the one in Tile calorimeter increases with the energy
increasing.

\section{Conclusions}
\hspace{6mm}
The experimental longitudinal hadronic shower profiles in combined
calori\-meter consisting of the lead-argon electromagnetic part and 
iron-scintilla\-tor hadronic part have been obtained.
The degree of description of the generally accepted Bock parameterization
of the longitudinal shower development has been investigated.
It is shown that this parameterization does not give satisfactory description
for this combined calorimeter.
Some modification of this parameterization, in which the $e/h$ ratios of
the compartments of combined  calorimeter are used, is suggested and compared 
with experiment.
The agreement between such parameterization and the experimental data
is demonstrated.

\section{Acknowledgments}
\hspace{6mm}
This work is the result of the efforts of many people from the ATLAS
Collaboration.
The authors are greatly indebted to all Collaboration for their test beam 
setup and data taking.
Authors are grateful Peter Jenni and Marzio Nessi for fruitful discussion
and support of this work. 
We are thankful Julian Budagov and Jemal Khubua for their attention and 
support of this work.
We are also thankful Illias Efthymiopoulos and Irene Vichou for 
fruitful discussion.




\vspace*{\fill}
\begin{table}[tbph]
\begin{center}
\caption{
	The differential energy depositions $\Delta E/ \Delta x$
	as a function of the longitudinal coordinate $x$ for the various 
	beam energies.
	} 
\label{T1}
\begin{tabular}{|c|c|c|c|c|c|}  
\multicolumn{6}{c}{\mbox{~~}}\\
\hline
    N&  x&\multicolumn{4}{|c|}{$E_{beam}$ (GeV)}\\
\cline{3-6}
depth&($\lambda_{\pi}$)&10  &20           &40	    &50 		\\ 
\hline
1  &0.294&$5.45\pm0.08$  & $8.58\pm0.16$ &$14.3\pm0.2$   &$16.6\pm0.4$ \\
2  &0.681&$4.70\pm0.08$  & $9.10\pm0.15$ &$16.7\pm0.2$   &$20.8\pm0.3$ \\
3  &1.026&$2.66\pm0.06$  & $5.55\pm0.11$ &$11.1\pm0.2$   &$13.6\pm0.2$ \\
dm &1.315&$1.35\pm0.07$  & $2.75\pm0.14$ &$5.28\pm0.26$  &$6.46\pm0.32$ \\
4  &2.06 &$1.93\pm0.03$  & $4.35\pm0.06$ &$8.99\pm0.08$  &$11.0\pm0.1$ \\
5  &3.47 &$0.87\pm0.02$  & $2.13\pm0.04$ &$5.29\pm0.06$  &$6.15\pm0.10$ \\
6  &5.28 &$0.18\pm0.01$  & $0.57\pm0.02$ &$1.50\pm0.03$  &$2.07\pm0.05$ \\
7  &7.50 &$0.025\pm0.003$& $0.11\pm0.01$ &$0.32\pm0.01$  &$0.49\pm0.02$\\
\hline
\hline
    N&  x&\multicolumn{4}{|c|@{}}{$E_{beam}$ (GeV)}\\
\cline{3-6}
depth&($\lambda_{\pi}$)&80&100&150&300 \\ 
\hline
1  &0.294&$22.6\pm0.6$ &$28.4\pm0.6$ &$36.3\pm0.7$ &$61.3\pm1.5$ 	\\
2  &0.681&$30.4\pm0.4$ &$37.6\pm0.5$ &$53.5\pm0.8$ &$97.9\pm1.7$ 	\\ 
3  &1.026&$20.3\pm0.3$ &$25.7\pm0.4$ &$37.2\pm0.6$ &$68.9\pm1.2$ 	\\ 
dm &1.315&$10.1\pm0.5$ &$12.8\pm0.6$ &$19.0\pm1.0$ &$34.1\pm1.7$   	\\   
4  &2.06 &$18.0\pm0.1$ &$22.4\pm0.2$ &$33.9\pm0.3$ &$64.8\pm0.7$ 	\\   
5  &3.47 &$11.9\pm0.1$ &$14.6\pm0.2$ &$23.3\pm0.2$ &$49.0\pm0.5$ 	\\
6  &5.28 &$3.66\pm0.06$&$4.57\pm0.08$&$8.18\pm0.13$&$18.6\pm0.3$ 	\\
7  &7.50 &$0.86\pm0.03$&$1.10\pm0.04$&$2.04\pm0.06$&$5.54\pm0.15$	\\   
\hline
\end{tabular}
\end{center}
\end{table}
\vspace*{\fill}


\newpage
\clearpage
\clearpage
\clearpage
\clearpage
\clearpage
\clearpage
\clearpage
\clearpage
\clearpage

\end{document}